\documentclass[12pt]{article}

\usepackage{amssymb, amsmath}
\usepackage[T1]{fontenc}
\usepackage{graphicx}
\usepackage{cite}
\usepackage{hyperref}
\usepackage{comment}
\usepackage[utf8]{inputenc} 
\def\be{\begin{equation}}
\def\ee{\end{equation}}
\def\bea{\begin{eqnarray}}
\def\eea{\end{eqnarray}}
\def\bes{\begin{equation*}}
\def\ees{\end{equation*}}
\def\beas{\begin{eqnarray*}}
\def\eeas{\end{eqnarray*}}

\def\td{ \textrm{d}}

\def\zo{^{(0)}}
\def\fo{^{(1)}}
\def\so{^{(2)}}
\def\no{^{(n)}}
\def\sgk{\kappa^{(\ell)}}

\def\nn{\nonumber}
\usepackage{xcolor}
\usepackage{leftidx}

\title{ {\bf Spacetime near Kerr isolated horizon}}

\author{Jerzy Lewandowski$^a$\footnote{Jerzy.Lewandowski@fuw.edu.pl}\; and Carmen Li$^{ab}$\footnote{k.k.li@lancaster.ac.uk}  
\\ \\  \small \sl  $^a$Faculty of Physics, University of Warsaw, ul. Pasteura 5, 02-093 Warsaw, Poland
\\ \small \sl $^b$ 
 School of Computing and Communications, InfoLab21,
 \\ \small \sl Lancaster University, LA1 4WA, Lancaster, U.K.
 }

\date{}

\begin{document}

\maketitle

\begin{abstract}
The theory of isolated horizon provides a quasi-local framework to study the spacetime geometry in the neighbourhood of the horizon of a black hole in equilibrium without any reference to structures far away from the horizon. While the geometric properties of the Kerr-(A)dS and more general algebraically special solutions have drawn substantial interest recently in the isolated horizon formalism, their horizon metrics have never been written down explicitly in the adapted Bondi-like coordinate system. Following the approach by Krishnan and assuming that the horizon symmetry extends to certain order in the bulk, we present in this note a general method to compute the metric functions order by order radially in Bondi-like coordinates in 4-dimensions from a small set of intrinsic data -- the connection and the Newman-Penrose spin coefficient $\pi$ specified on the horizon cross-section. Applying this general method, we then present the horizon metric of non-extremal Kerr-dS in Bondi-like coordinates. For the pure Kerr case without a cosmological constant, we also show explicitly the metric functions to the first order. 
\end{abstract}

\newpage

\section{Introduction}
The notion of black hole event horizon requires the knowledge of the global structure of the entire spacetime; however, it is not possible to access the whole spacetime all the way to \textit{scri} in any physically viable situation. On the other hand, it is interesting to investigate the spacetime in the near-horizon region alone of a black hole, regardless of any structures further away. This is especially true in astrophysical context, where one may want to focus just on an isolated systems of a single or multiple black holes and nothing else. The theory of isolated horizon\cite{Ashtekar:2001jb} provides such a framework to probe the spacetime geometry in the neighbourhood of the black hole horizon; it describes quasi-locally a black hole in equilibrium with its surroundings in a sense the horizon is \textit{non-expanding}, with no reference to anything beyond a small neighbourhood of the horizon. While an isolated horizon admits symmetry generated by null vector field on the horizon, unlike a stationary black hole spacetime whose event horizon is a Killing horizon of a globally defined Killing field which is asymptotically timelike, the horizon symmetry on an isolated horizon may \textit{not} extend into the bulk outside the horizon. The isolated horizon framework therefore allows one to study black hole spacetimes with fewer, weaker and less restrictive symmetry assumptions. 
\\
\\
The isolated horizon formalism has been used for instance to prove several \textit{local} black hole uniqueness theorems \cite{Lewandowski:2000nh}\cite{Lewandowski:2002ua} \cite{Dobkowski-Rylko:2018ahh}, and remarkably, to prove by assuming Petrov type D and bifurcated horizon, the existence of axisymmetry in Kerr horizon without rigidity \cite{Lewandowski:2018rau}. Unlike the conventional ``black hole uniqueness'' for stationary black holes (see e.g. \cite{Chrusciel:2012jk} for review), no assumptions have been made regarding stationarity, asymptotic flatness, global hyperbolicity nor analyticity of the spacetime. Higher dimensional black holes are well known to violate black hole uniqueness, even in vacuum: there are the Myers-Perry and Emparan-Reall black ring solutions in 5d which are both stationary and asymptotically flat, but their different horizon topologies cannot be distinguish by any asymptotic Gaussian flux integral. Of course finding new, analytic balck hole solutions to the Einstein equations is not an easy task; one may instead focus on the simpler problem of finding the possible \textit{horizon} topology and geometry. Some interesting results in the context of \textit{local} characterisation of black hole horizons have also been found recently within the isolated horizon framework \cite{Dobkowski-Rylko:2018usb}, which may play an important role in the overall black hole \textit{classification} problem. 
\\
\\
In order to make use of the isolated horizon formalism, one needs to work in Bondi-like or Gaussian null coordinates, which are adapted coordinates regular on the horizon. Nevertheless, there are no examples of even well-known solutions written explicitly in such coordinates in the literature. A recent attempt to put the Kerr-Newman solution into Bondi-like coordinates by transforming the Kinnersley tetrad runs into an integral that cannot be evaluated explicitly \cite{Scholtz:2017ttf}. In this note, we will present a general method to compute the spacetime metric of any arbitrary solution admitting an isolated horizon in the neighbourhood of the horizon in Bondi-like coordinate system, \textit{order by order in the null radial direction} $r$, by following the approach introduced in \cite{Krishnan:2012bt}, assuming that \textit{the horizon symmetry extends beyond the horizon} to certain orders in $r$. Although our result is only valid in a small neighbourhood of the horizon, it is still useful in for example, astrophysical context where the background metric in the vicinity of a back hole is required. \\
\\
By virtue of the characteristic initial value problem \cite{Friedrich:1981at}\cite{Rendall}, and more recently ``black hole holograph'' \cite{Racz:2013hva}, assuming that the spacetime admits a pair of intersecting null hypersurfaces generated by expansion and shear free geodesically compete null congruences in Einstein vacuum theory, the spacetime geometry is uniquely determined by a small set of data specified on the space-like intersection, namely the complex vector field which determines the induced metric on the intersection, as well as the Newman-Penrose spin coefficient $\pi$. Thus given these pieces of data on the horizon cross-section, we can reconstruct the full spacetime in Bondi-like form by systematically applying the Newman-Penrose equations and Bianchi identities. Since we will be exploiting the full advantage of the desirable properties arising from the Newman-Penrose formalism, we will restrict our discussion to 4(3+1) spacetime dimensions.  \\
\\
Our analysis holds for any \textit{non-extremal} Kerr(-dS) isolated horizon. As we shall from the equations, this method cannot be applied to extremal black holes due to divergence at extremality. The analogous problem for extremal black holes has been investigated in \cite{Li:2015wsa} under the framework of \textit{near-horizon geometry}.

\section{Preliminaries}
\subsection{Notations}
Although our conventions follow closely those in \cite{Krishnan:2012bt}, for completeness and explicitness, we shall include a section to clarify our notations. Our spacetime metric has signature $(-,+,+,+)$; the 4-spacetime indices are denoted by Greek letter $\alpha, \beta,...,\mu, \nu,...$ (not to be confused with the Newman-Penrose coefficients); 3-horizon indices are denoted by lower case Latin letters $a,b,...$ and the 2-spatial cross-section indices are denoted by upper case Latin letters $A,B,...$. 
\subsection{Isolated horizon}
Let us begin with the definition of an isolated horizon and the introduction of an adapted, regular coordinate system near the horizon. We shall not give a full account of the theory of isolated horizon (IH) here but rather state the important properties which are relevant for our purpose. \\
\\
Consider a smooth co-dimension 1 null surface $\Delta$ embedded in the 4-dimensional spacetime $(M, g)$, which for our case of interest would be horizon of a black hole in equilibrium. Let $\ell$ be a null normal of $\Delta$, then it is geodesic and the surface gravity associated with $\ell$ is defined as its self-acceleration  
\bes
\ell^a  \nabla_a \ell^b = \kappa^{(\ell)}  \ell^b \; . 
\ees
Extremal horizons are those with vanishing $\kappa^{(\ell)}$; otherwise they are non-extremal. Let $q_{ab}$ be the degenerate metric with signature $(0,+,+)$ induced on $\Delta$ from the spacetime metric $g_{\mu _\nu}$. Assuming that Einstein field equations hold on $\Delta$, then it is a \textit{non-expanding horizon} if:\\
\indent (i) the expansion of any null normal $\ell$ vanishes, i.e. 
\bes
\theta_\ell \equiv q^{ab} \nabla_a \ell_b =0
\ees
\indent where $q^{ab}$ is an inverse of the induced metric $q_{ab}$ (note that because $q_{ab}$ is degenerate, $q^{ab}$ is not unique) and $\nabla_a$ is the covariant derivative induced on $\Delta$.\\
\indent (ii) $\Delta$ has topology $S^2 \times \mathbb{R}$ . \\
\indent (iii) The energy condition that the stress energy tensor $T_{ab}$ satisfies $-T^a_{~b} \ell^b$ is causal and future directed on $\Delta$. \\
It can be shown that as a result of the above conditions, the degenerate induced metric is Lie dragged by $\ell^a$, i.e.
\bes
\mathcal{L}_\ell q^{ab} =0 \; , 
\ees
so $\ell$ is a Killing vector \textit{on} $\Delta$. However, unlike a Killing horizon which is defined as a null hypersurface in spacetime on which a Killing vector field of the spacetime becomes null, there needs not exist \textit{any} Killing field in the neighbourhood of $\Delta$. 
A non-expanding horizon is in particular, an \textit{isolated horizon} if in addition 
\bes
[\mathcal{L}_\ell , \nabla_a]=0 \; . 
\ees
\subsection{Coordinate system}
We can now introduce a coordinate system adapted to $\Delta$ in its neighbourhood, which is regular on $\Delta$. Such coordinate system is often called Gaussian null coordinates or Bondi-like coordinates in the literature. Without loss of generality, $\ell$ can be chosen to be future directed and we may assign $v$ as an affine parameter along integral curves of $\ell$ on $\Delta$ and make it a coordinate. $\Delta$ is thus foliated by $S_v$, spatial sphere\footnote{The spatial cross-section does not have to be topologically $S^2$ in general.} labelled by $v$, and we can then assign local coordinates $x^A$ on $S_0$. Starting from a point $p \in S_0$, a point $q \in \Delta$ lying an affine parameter $v$ away along the integral curve of $\ell$ is therefore labelled by coordinates $(v, x^A)$, by keeping the functions $x^A$ constant along the curve. The coordinate $v$ is a null-time coordinate on $\Delta$. 
\\
\\
Now we need to extend the coordinate chart into the bulk. At every point on $\Delta$, let $n$ be the unique future directed null vector field satisfying the normalisation $\tfrac{\partial}{\partial v} \cdot n = -1$ and orthogonality $n \cdot \tfrac{\partial}{\partial x^A} =0$. Starting from point $q$, a point $s \in M$ lying an affine parameter value $r$ away along the null geodesic with tangent $-n$ on $\Delta$ is then assigned the coordinates $(v, r, x^A)$, by keeping the functions $v$ and $x^A$ constant along the geodesic. Thus $(v, r, x^A)$ define a coordinate chart in the neighbourhood of $\Delta$ via a double foliation with the horizon $\Delta$ located at $r=0$. This coordinate chart is valid as long as the null geodesics do not develop caustics.  \\
\\
We have on the horizon $\Delta$, a pair of future-directed null vectors $\left. n \right|_\Delta = - \tfrac{\partial}{\partial r}$ and $\left. \ell \right|_\Delta = \tfrac{\partial}{\partial v}$. How do $n$ and $\ell$ extend to the \textit{bulk}? By construction, the integral curves of $\tfrac{\partial}{\partial r}$ are null geodesics so $\tfrac{\partial}{\partial r}$ is null everywhere, we can simply identify $n= - \tfrac{\partial}{\partial r}$. However, $\tfrac{\partial}{\partial v}$ is not null in the bulk in general. Hence the null vector $\ell =  \tfrac{\partial}{\partial v} + U \tfrac{\partial}{\partial r} + X^A \tfrac{\partial}{\partial x^A}$ with $U$ and $X^A$ vanishing on the horizon. \\
\\
How about the normalisation and orthogonality conditions? It can be shown that indeed, both $\nabla_n \left( n \cdot \tfrac{\partial}{\partial v} \right) =0$ and $\nabla_n \left( n \cdot \tfrac{\partial}{\partial x^A} \right) =0$, therefore the metric components $g_{vr}= n \cdot \tfrac{\partial}{\partial v} =1$ and $g_{rA} = n \cdot \tfrac{\partial}{\partial x^A} = 0 $ everywhere, not just on $\Delta$.  \\
\\
Nevertheless, $g_{va}$ need not vanish outside $\Delta$, nor does $g_{vv}$ as we already stated earlier. We now have all the ingredients to write down the spacetime metric in Gaussian null coordinates. In the neighbourhood of in fact, any null hypersurface $\mathcal{N}$, it takes the form 
\bes
g = 2 \td v \left( \tfrac{1}{2} r f \td v + \td r + r h_A \td x^A  \right) + \gamma_{AB} \td x^A \td x^B
\ees
where $f$, $h_A$ and $\gamma_{AB}$ are in general, smooth functions of all coordinates $(v, r, x^A)$ .

\subsection{The Newman-Penrose formalism}
In order to make use of the characteristic initial value formulation to obtain the spacetime in the neighbourhood of a Kerr IH, we shall work in the Newman-Penrose (NP) tetrad formalism. The NP tetrad consist of two real null vectors $\ell$ and $n$, and two complex null vectors $m$ and $\bar{m}$ which are conjugates of each other. They satisfy the cross-normalisations $\ell \cdot n = -1$ and $m \cdot \bar{m} =1$ with all other inner products vanishing. Thus the metric is given by 
\bes
g_{\mu \nu} = - \ell_\mu n_\nu - n_\mu \ell_\nu + m_\mu \bar{m}_\nu +\bar{m}_\mu m_\nu
\ees
We already have candidates for $\ell$ and $n$ from above; we may choose $m$ on the horizon to be a complex null vector on $S_0$ and Lie drag it along $\ell$ i.e. $\mathcal{L}_\ell m^a =0$ on $\Delta$. Along with the inner product rules, we deduce that the null vector $m$ must be of the form 
\bes
m = \Omega \frac{\partial}{\partial r} + \xi^A \frac{\partial}{\partial x^A} \; , 
\ees
where the functions $\Omega$ and $\xi^A$ are complex and $\xi^A$ satisfy the inner product rules $\xi \cdot \xi |_\Delta= \bar{\xi} \cdot \bar{\xi}|_\Delta =0$ and $\xi \cdot \bar{\xi}|_\Delta =1$ on the horizon $\Delta$. It is clear that $m$ should have no $\tfrac{\partial}{\partial v}$ component because of the condition $m \cdot n =0$. Since we want $m$ to be tangent to spheres $S_v$ on $\Delta$, $\Omega$ must vanish on $\Delta$. \\
\\
In summary, the null vector (and co-vector) fields that define the null frame can be expressed as follows:\\
As vector fields
\beas
\ell^\mu \partial_\mu &=& \frac{\partial}{\partial v} + U \frac{\partial}{\partial r} + X^A \frac{\partial}{\partial x^A} \\
n^\mu \partial_\mu &=& - \frac{\partial}{\partial r} \\
m^\mu \partial_\mu &=& \xi^A \left( \frac{\partial}{\partial x^A} - Z_A \frac{\partial}{\partial r} \right) \; ,
\eeas
where our previously defined $\Omega := \xi^A Z_A $ and as 1-forms
\beas
\ell_\mu \td x^\mu &=& \td r + Z_A \td x^A - \left( U + Z_A X^A \right) \td v \\
n_\mu \td x^\mu &=& - \td v \\
m_\mu \td x^\mu &=& \xi_A \left( \td x^A - X^A \td v \right) \; 
\eeas
where $U$, $X^A$ and $Z_A$ are real functions and vanish on $\Delta$. The metric functions are related to $(U, X^A,Z_A, \xi)$ by 
\beas
g_{vv}&=& - (U + Z_A X^A) \\
g_{vr}&=& 1 \\
g_{vA}&=& Z_A - \xi_A \bar{\xi}_B X^B - \bar{\xi}_A {\xi}_B X^B  \\
g_{AB}&=&  \xi_A \bar{\xi}_B + \bar{\xi}_A {\xi}_B  
\eeas
and the rest are zero by construction. \\
\\
We will also denote the directional derivatives associated with the null vectors by 
\bes
D:= \ell^\mu \nabla_\mu \; , \quad \Delta:= n^\mu \nabla_\mu \; ,  \quad \delta:= m^\mu \nabla_\mu \;, \quad \mathrm{and}  \quad \bar{\delta}:= \bar{m}^\mu \nabla_\mu \; . 
\ees
The null tetrad, which are typically non-coordinate basis, satisfy the following commutation relations for any function $f$, in terms of \textit{Newman-Penrose spin coefficients}:
\bea
\nn  \left( \Delta D - D \Delta \right) f &=& (\epsilon + \bar{\epsilon}) \Delta f + (\gamma + \bar{\gamma}) D f - (\bar{\tau} + \pi ) \delta f - (\tau + \bar{\pi}) \bar{\delta} f \\ 
\nn  \left( \delta D - D \delta \right) f &=& (\bar{\alpha} + \beta - \bar{\pi} ) D f + \kappa \Delta f - (\bar{\rho} + \epsilon - \bar{\epsilon}) \delta f - \sigma \bar{\delta} f \\
\nn  \left( \delta \Delta -  \Delta \delta \right) f &=& - \bar{\nu} D f + (\tau - \bar{\alpha} - \beta) \Delta f + (\mu - \gamma + \bar{\gamma} ) \delta f + \bar{\lambda} \bar{\delta} f \\
\left(\bar{\delta} \delta  - \delta \bar{\delta} \right) f &=&  (\bar{\mu} - \mu ) D f + (\bar{\rho} - \rho) \Delta f + (\alpha - \bar{\beta}) \delta f - (\bar{\alpha} - \beta) \bar{\delta} f \label{comrel} \; .  
\eea
\subsection{Radial expansion}
Since our goal is to find the spacetime structure in the neighbourhood of the Kerr(-dS) isolated horizon, we shall expand the the directional derivatives, NP spin coefficients and Weyl tensor components in the radial direction $r$ away from the horizon. Thus we may write any quantity $X$ as 
\bes
X = X^{(0)} + r X^{(1)} + \tfrac{1}{2} r^2 X^{(2)} +...
\ees
and similarly for the directional derivatives, e.g. 
\beas
\delta X &=&  \delta^{(0)} X^{(0)} +  r \left(  \delta^{(1)} X^{(0)}  + \delta^{(0)} X^{(1)}  \right) +...\\
&=&  \xi^{A(0)} \frac{\partial X^{(0)}}{\partial x^A}+ r \left( \xi^{(1)} \frac{\partial X^{(0)}}{\partial x^A} + Z_a^{(1)} \frac{\partial X^{(0)}}{\partial r} +\xi^{(0)} \frac{\partial X^{(1)}}{\partial x^A} + Z_a^{(0)} \frac{\partial X^{(1)}}{\partial r}      \right) 
+ ...
\eeas
for some arbitrary complex function $X$. 
\section{Field equations for spacetime admitting an isolated horizon}
As we shall see below, the field equations can be divided into three categories: radial equations which contain the radial derivatives $\Delta$, \textit{evolution equations} which contain derivatives along the null-time direction $v$ i.e. $D$ but not $\Delta$, and \textit{angular equations} which involve only the $\delta$ and $\bar{\delta}$. Note that they are named after the properties of the derivatives in leading order or equivalently, on the horizon; off the horizon e.g. the derivative $D$ does involve the radial derivative $\tfrac{\partial}{\partial r}$. \\
\\
Since for our spacetime containing an isolated horizon the integral curves of $n$ are affinely parametrised geodesics and $\ell$, $m$ are parallelly propagated along $n$, $\Delta n = \Delta \ell = \Delta m=0$, which means $\gamma = \tau = \nu =0$. Furthermore, by considering $f=v$ in the commutation relations, it is straight forward to deduce that $\pi= \alpha + \bar{\beta} $ and $\mu= \bar{\mu} \in \mathbb{R}$. Therefore the Newman-Penrose equations simply as follows:
\subsection*{Radial equations}
\beas
0 &=& \Delta \kappa + \bar{\pi} \rho + \pi \sigma + \Psi_1 \\
0 &=& \Delta \epsilon + \bar{\pi} \alpha + \pi \beta + \Psi_2 - \tfrac{R}{24}\\
0 &=& \Delta \pi + \pi \mu + \bar{\pi} \lambda + \Psi_3\\
0 &=& \Delta \lambda + 2 \mu \lambda + \Psi_4\\
0 &=& \Delta \mu + \mu^2 + |\lambda|^2 \\
0 &=& \Delta \beta + \mu \beta + \alpha \bar{\lambda} \\
0 &=& \Delta \sigma + \mu  \sigma + \bar{\lambda} \rho \\
0 &=& \Delta \rho + \rho \mu + \sigma \lambda + \Psi_2 + \tfrac{R}{12}\\
0 &=& \Delta \alpha + \alpha \mu + \beta \lambda + \Psi_3
\eeas
\subsection*{Time evolution equations}
\beas
D \rho - \bar{\delta} \kappa &=& \rho^2 + \sigma \bar{\sigma} + (\epsilon + \bar{\epsilon})\rho - 2 \alpha \kappa \\
D \sigma - \delta \kappa &=& ( \rho + \bar{\rho}) \sigma + (3 \epsilon - \bar{\epsilon})\sigma - 2 \beta \kappa + \Psi_0 \\
D \alpha - \bar{\delta} \epsilon &=& \alpha (\rho + \bar{\epsilon} - 2 \epsilon) + \beta \bar{\sigma} - \bar{\beta} \epsilon - \kappa \lambda + (\epsilon + \rho)\pi  \\
D \beta - \delta \epsilon &=&  (2 \alpha + \bar{\beta}) \sigma + (\bar{\rho} - \bar{\epsilon}) \beta - \mu \kappa + \beta \epsilon + \Psi_1\\
D \lambda - \bar{\delta} \pi &=& (\rho - 3 \epsilon + \bar{\epsilon}) \lambda + \bar{\sigma}\mu + 2 \alpha \pi \\
D \mu - \delta \pi &=& (\bar{\rho} - \epsilon - \bar{\epsilon}) \mu  + \sigma \lambda + 2 \beta \pi + \Psi_2
\eeas
\subsection*{Angular field equations}
\beas
\delta \rho - \bar{\delta} \sigma &=& \bar{\pi} \rho - \sigma (3 \alpha - \bar{\beta}) - \Psi_1 \\
\delta \alpha - \bar{\delta} \beta &=& \mu \rho - \lambda \sigma + |\alpha|^2 + |\beta|^2 - 2 \alpha \beta - \Psi_2 + \tfrac{R}{24}\\
\delta \lambda - \bar{\delta} \mu &=& \pi \mu + \lambda (\bar{\alpha} - 3 \beta)  \sigma (3 \alpha - \bar{\beta}) - \Psi_3
\eeas
Apart from the Newman-Penrose equations, the Weyl tensor components also satisfy the Bianchi identities:
\subsection*{Radial Bianchi identities}
\beas
\Delta \Psi_0 - \delta \Psi_1 &=& - \mu \Psi_0  - 2 \beta \Psi_1 + 3 \sigma \Psi_2 \\
\Delta \Psi_1 - \delta \Psi_2 &=& - 2 \mu \Psi_1  + 2  \sigma \Psi_3 \\
\Delta \Psi_2 - \delta \Psi_3 &=&  \sigma \Psi_4 + 2 \beta - 3 \mu \Psi_2 \\
\Delta \Psi_3 - \delta \Psi_4 &=& 4 \beta \Psi_4 - 4 \mu \Psi_3 
\eeas
\subsection*{Time evolution Bianchi identities}
\beas
D \Psi_1 - \bar{\delta} \Psi_0 &=& (\beta - 3 \alpha) \Psi_0 + 2 (2 \rho + \epsilon) \Psi_1 - 3 \kappa \Psi_2 \\
D \Psi_2 - \bar{\delta} \Psi_1 &=& -\lambda \Psi_0 + 2 \bar{\beta} \Psi_1 + 3 \rho \Psi_2 - 2 \kappa \Psi_3 \\
D \Psi_3 - \bar{\delta} \Psi_2 &=& - 2 \lambda \Psi_1 + 3 \pi \Psi_2  + 2 (\rho - \epsilon ) \Psi_3 - \kappa \Psi_4 \\
D \Psi_4 - \bar{\delta} \Psi_3 &=& - 3 \lambda \Psi_2 + 2 (3 \alpha + 2 \bar{\beta} ) \Psi_3 - (4 \epsilon - \rho) \Psi_4
\eeas
\subsection*{Metric functions} \label{LOmetricfn}
The real functions $U$, $X^A$ and $Z_A$ and the complex function $\xi^A$ satisfy the following equations from the commutation relations \eqref{comrel}. They are obtained by replacing the arbitrary function $f$ with coordinates $r$ and $x^A$, and we make use of what we have obtained above to simplify them; the equations from the coordinate $v$ don't give any new information. These eight equations can also be divided into radial, evolutionary and angular equations. The radial equations are
\bea
\nn \Delta U &=& - \kappa^{(\ell)} - \bar{\pi} \bar{\Omega} - \pi \Omega \\
\nn \Delta X^A &=&  - \bar{\pi} \bar{\xi}^A - \pi \xi^A \\
\nn \Delta \xi^A &=& - \mu \xi^A - \bar{\lambda} \bar{\xi}^A \\
\xi^A \Delta Z_A &=& - \bar{\pi} - 2 \mu \xi^A Z_A - 2 \bar{\lambda} \bar{\xi}^A Z_A \; , \label{radmetricfn}
\eea
the last equation is complex hence it really is two real equations for the two real functions $Z_1$ and $Z_2$. The evolutionary equations are
\beas
D \left( \xi^A Z_A \right) - \delta U &=& \kappa + \rho  \xi^A Z_A  + \sigma  \bar{\xi}^A Z_A \\
D \xi^A - \delta X^A &=& \bar{\rho} \xi^A + \sigma \bar{\xi}^A \; . 
\eeas
Finally, the angular equations are 
\beas
\bar{\delta} \xi^A - \delta \bar{\xi}^A &=& (\alpha - \bar{\beta} ) \xi^A - (\bar{\alpha} - \beta) \bar{\xi}^A  \\
\left( \xi^A \bar{\delta} - \bar{\xi}^A \delta \right) Z_A &=& \rho - \bar{\rho} \; . 
\eeas

\section{Independent horizon data} \label{indepHD}
In this section we examine the horizon data. We write down the NP equations at $\mathcal{O}(r^0)$. Clearly, not all nine of the remaining NP coefficients are independent, nor are the Weyl tensor components. Recall that $\ell$ is geodesic on $\Delta$, expansion free and orthogonal to the spheres $S_v$. This yields $\rho^{(0)} = \kappa^{(0)} =0$. We may also choose $\xi^{(0)}$ such that $ \epsilon^{(0)} \in \mathbb{R} $. Moreover, the spin 1-form on $\Delta$ is written in NP coefficient as 
\bes
\omega_a = (\epsilon + \bar{\epsilon} ) n_a + \pi^{(0)} m_a + \bar{\pi}^{(0)} \bar{m}_a \; , 
\ees
$\Delta$ being an isolated horizon means $\mathcal{L}_\ell \omega_a =0$, which yields
\bes
\kappa^{(\ell)} = \epsilon + \bar{\epsilon} = 2 \epsilon =\mathrm{constant} \; , \quad D\pi^{(0)} = 0 \; . 
\ees
Let us first look at what these imply to the time evolution equations on an isolated horizon $\Delta$. The first two give $\sigma^{(0)} =0$ and $\Psi_0^{(0)}$. The rest of them give
\bea
\nn && \frac{\partial }{\partial v} \alpha^{(0)} =0 \\
\nn && \frac{\partial }{\partial v} \beta^{(0)} =0 \\
&& \left( \frac{\partial }{\partial v} + \kappa^{(\ell)} \right)  \lambda^{(0)} = \left( \bar{\xi}^{A(0)} \frac{\partial}{\partial x^A} + 2 \alpha^{(0)} \right) \pi^{(0)} \label{evolutionlambda}\\
&& \left( \frac{\partial }{\partial v} + \kappa^{(\ell)} \right)  \mu^{(0)} = \left( {\xi}^{A(0)}  \frac{\partial}{\partial x^A} + 2 \beta^{(0)} \right) \pi^{(0)} + \Psi_2^{(0)} + \frac{R}{12} \label{evolutionmu} \; . 
\eea
The latter two can be integrated to give time evolutions of $\mu^{(0)}(v)$ and $\lambda^{(0)}(v)$
\beas
\mu^{(0)} &=& \mu^{(0)}(0) \exp \left(- \kappa^{(\ell)} v \right)+ \frac{1}{\kappa^{(\ell)}} \left( {\xi}^{A(0)}  \frac{\partial}{\partial x^A}  \pi^{(0)}  + 2 \beta^{(0)}  \pi^{(0)} + \Psi_2^{(0)} + \frac{R}{12} \right) \left( 1- \exp \left(- \kappa^{(\ell)} v \right)\right)\\
\lambda^{(0)} &=& \lambda^{(0)}(0) \exp \left(- \kappa^{(\ell)} v \right)+ \frac{1}{\kappa^{(\ell)}} \left( \bar{\xi}^{A(0)}  \frac{\partial}{\partial x^A}  \pi^{(0)}  + 2 \alpha^{(0)}  \pi^{(0)} \right) \left( 1- \exp \left(- \kappa^{(\ell)} v \right)\right)
\eeas
The first angular field equation gives simply 
\bes
\Psi_1^{(0)} =0 \; ;
\ees
the second equation gives 
\beas
\mathrm{Re} \left[\Psi_2^{(0)} \right]&=& - \frac{\leftidx{^2} R}{4}  + \frac{R}{24} \\
i \mathrm{Im} \left[ \Psi_2^{(0)} \right] &=& - \frac{1}{2} \left( \eth \pi^{(0)} - \bar{\eth} \bar{\pi}^{(0)} \right) \; , 
\eeas
where $\leftidx{^2} R$ is the Ricci scalar on $S_0$, $\eth$ and $\bar{\eth}$ are spin raising and lowering operators as given in \cite{Krishnan:2012bt}. The last angular equation gives
\bes
\Psi_3^{(0)} = \left(\bar{\delta}^{(0)} + \pi\zo \right) \mu\zo + \left(-\delta\zo + \bar{\alpha}\zo - 3 \beta\zo \right) \lambda\zo \; . 
\ees
Finally, let us turn to the evolutionary Bianchi identities. The first one is trivial following from what we have found already, whereas the second one yields
\bes
\frac{\partial}{\partial v} \Psi_2\zo =0  \; . 
\ees 
The third one can also be showed to be redundant given the above. The last one gives an evolution equation for $\Psi_4\zo$
\bes
\left( \frac{\partial}{\partial v} + 2 \sgk \right) \Psi_4\zo = \left( \bar{\delta}  + 6 \alpha\zo + 4 \bar{\beta}\zo \right) \Psi_3\zo - 3 \lambda\zo \Psi_2\zo  \; , 
\ees
which again can be integrated with $\Psi_4\zo(v=0)$ a free datum on $S_0$. \\
\\
Therefore on any isolated horizon $\Delta$, after choosing a cross section $S_0$ on $\Delta$ and fixing the horizon generator $\ell^a$ along with its surface gravity $\kappa^{(\ell)}$, only the following data can be chosen independently on $S_0$: the connection\footnote{It is called a connection because $\alpha\zo - \bar{\beta}\zo = \frac{1}{2} \left( \bar{\xi}^{(0)A} \bar{\xi}^{(0)B} \nabla_B \xi\zo_A - {\xi}^{(0)A} \bar{\xi}^{(0)B} \nabla_B \bar{\xi}\zo_A  \right)$, where $\nabla_A$ is the covariant derivative with respect to the horizon cross-section metric $\gamma$. } $A \equiv \alpha\zo - \bar{\beta}\zo $ and $\pi\zo = \alpha\zo + \bar{\beta}\zo$ which are time $(v)$ independent on $\Delta$, and the transversal expansion and shear $\lambda\zo$ and $\mu\zo$, as well as the Weyl component $\Psi_4\zo$, which evolve in time on $\Delta$ according to the expressions given above. This is the same as the result found in \cite{Krishnan:2012bt}, despite the fact that we have included a cosmological constant ($R\neq 0$). 
\section{First order expansion}
Now we may use the radial NP equations and the radial Bianchi identities to obtain \textit{algebraically} the nine non-trivial NP coefficients and the Weyl components $\Psi_i$, $i=0,1,2,3$ from the lower order data. Note that there is \textit{no} radial equation for $\Psi_4$. In fact, in order to employ the characteristic initial value problem, $\Psi_4|_{v=0} =\Psi_4\zo (x) + r \Psi_4\fo  (x) + \tfrac{1}{2} r^2 \Psi_4^{(2)}  (x) +... $ is also a free datum we need to specify on $\mathcal{N}_0 = {v=0}$, the transversal null surface to $\Delta$ intersecting at $S_0$. Explicitly, the first order data is given by
\bea
\nn \kappa\fo &=& 0 \\
\nn \epsilon\fo &=& \bar{\pi}\zo \alpha\zo + \pi\zo \beta\zo + \Psi_2\zo - \frac{R}{24} \\
\nn \pi\fo &=& \pi\zo \mu\zo + \bar{\pi}\zo \lambda\zo + \Psi_3\zo\\
\nn \lambda\fo &=& 2 \mu\zo \lambda\zo + \Psi_4\zo \\
\nn \mu\fo &=& \mu^{(0)2} + \left| \lambda\zo \right|^2 \\
\nn \beta\fo &=& \mu\zo \beta\zo + \alpha\zo \bar{\lambda}\zo \\
\nn \sigma\fo &=& 0 \\
\nn \rho\fo &=& \Psi_2\zo + \frac{R}{12} \\
\nn \alpha\fo &=& \beta\zo \lambda\zo + \mu\zo \alpha\zo + \Psi_3\zo \\
\nn \Psi_0\fo &=& 0 \\
\nn \Psi_1\fo &=&  - \delta\zo \Psi_2\zo \\
\nn \Psi_2\fo &=& 3 \mu\zo \Psi_2\zo - 2 \beta\zo \Psi_3\zo - \delta\zo \Psi_3\zo \\
\Psi_3\fo &=& - \delta\zo \Psi_4 - 4 \beta\zo \Psi_4\zo + 4 \mu\zo \Psi_3\zo   \label{radialfoeqn}
\eea

\subsection{Constructing the metric}
We need to compute the functions $U$, $X^A$, $Z_A$ and $\xi^A$ in order to obtain the metric. Recall that the $U$, $X^A$ and $Z_A$ are real and vanish on the horizon, and $\xi^{(0)A}$ are the complex null frame fields \textit{we choose} on the sphere $S_0$. Because of \eqref{radmetricfn}, it is clear that all frame fields can be obtained order by order from lower order data. For instance, to lowest order, the real functions
\beas
U\fo &=&  \kappa^{(\ell)} \\
X^{(1)A} &=&  \pi\zo  \xi^{(0)A} + \bar{\pi}\zo \bar{\xi}^{(0)A} \\
\xi^{(0)A} Z_A\fo &=&  \bar{\pi}\zo  \; , 
\eeas
and to first order, the complex function
\bes
\xi^{(1)A} =  \mu\zo \xi^{(0)A} + \bar{\lambda}\zo \bar{\xi}^{(1)A} \; . 
\ees
\section{Symmetry generated by $\frac{\partial}{\partial v}$ to $\mathcal{O}(r^n)$}
Now let us consider the special case where $\mathcal{L}_{\partial_v} g_{\mu \nu}^{(m)} =0$ for $0 \leq m \leq n$. At zeroth order, the differential operator $D\zo$ is trivial. In fact, there is no more ``evolutionary'' equations up to the $n$-th order as only they contain the derivatives $\tfrac{\partial}{\partial v}$. While most of the evolutionary equations above become redundant in the presence of this symmetry, three of them give us new information about the otherwise free data $\lambda\zo$, $\mu\zo$ and $\Psi_4$. \\
\\
The evolutionary equations for $\lambda$ \eqref{evolutionlambda} and $\mu$ \eqref{evolutionmu} become 
\beas
&&  \lambda^{(0)} = \frac{1}{ \kappa^{(\ell)}}\left( \bar{\xi}^{A(0)} \frac{\partial}{\partial x^A} + 2 \alpha^{(0)} \right) \pi^{(0)} \\
&&  \mu^{(0)} =  \frac{1}{ \kappa^{(\ell)}}\left[ \left( {\xi}^{A(0)}  \frac{\partial}{\partial x^A} + 2 \beta^{(0)} \right) \pi^{(0)} + \Psi_2^{(0)} + \frac{R}{12}  \right]\; ,
\eeas
therefore $\lambda\zo$ and $\mu\zo$ are \textit{no longer free} data on $S_0$. \\
\\
The evolutionary equations for $\Psi_4$ on the other hand becomes the radial equation for $\Psi_4$ up to order $n$, which gives 
\beas
\Psi_4\no &=& \frac{1}{\kappa^{(\ell)}}\left[ \bar{\Omega}\fo \Psi_3\no  + 2 \left(3 \alpha\zo + 2 \bar{\beta}\zo \right) \Psi_3\no + \bar{\Omega}\no \Psi_3\fo + \bar{\xi}^{(n)A} \frac{\partial \Psi_3\zo}{\partial x^A}  + 2 \left(3 \alpha\no + 2 \bar{\beta}\no \right) \Psi_3\zo  \right. \\
&&  \left. - 3 \lambda\zo \Psi_2\no - 3 \lambda\no \Psi_2\zo - U\no \Psi_4\fo - X^{(n)A} \frac{\partial \Psi_4\zo}{\partial x^A} - 4 \left( \epsilon\no - \rho\no \right) \Psi_4\zo  + \mathcal{O}(n-1)\right] \; ,
\eeas
with the $0$-th order equation giving its initial value  $\Psi_4\zo$ on the horizon in terms of other horizon data
\beas
\Psi_4\zo &=& \frac{1}{2 \kappa^{(\ell)}} \left(\bar{\delta}\zo  \Psi_3\zo + 2 \left(3 \alpha\zo + 2 \bar{\beta}\zo \right) \Psi_3\zo - 3 \lambda\zo \Psi_2\zo \right) \; .  
\eeas
Hence $\Psi_4$ is also no longer free data on the null surface $\mathcal{N}_0$. \\
\\
\section{Metric near Kerr isolated horizon}
In this section we compute explicitly the metric components in GNC near the Kerr isolated horizon. To begin with, we need to define the frame fields $\xi\zo$ on the horizon cross-section and $\Psi_a\zo$. Recall that in Boyer-Lindquist coordinates, the Kerr metric is given by 
\bes
\td s^2 = -\frac{\Delta_r}{\rho^2} \left( \td t - a \sin^2 \theta \td \phi \right)^2 + \frac{\rho^2}{\Delta_r} \td r^2 + \rho^2 \td \theta^2 + \frac{\sin^2 \theta}{\rho^2} \left( a \td t - (r^2+a^2) \td \phi \right)^2 
\ees 
where 
\bes
\rho^2 = r^2 + a^2 \cos^2 \theta \; ,  \quad \Delta_r = r^2 + a^2 - 2mr \; ,
\ees
$m$ is the mass and $a$ is the angular momentum. The event horizon is located at the $r= r_+$, the larger root of $ \Delta_r$, which is a Killing horizon generated by the Killing field
\bes
K = \frac{\partial}{\partial t} + \left(\frac{a}{r_+^2 + a^2} \right)\frac{\partial}{\partial \phi}\; , 
\ees
and the surface gravity is defined by 
\bes
\tilde{\kappa} = \sqrt{\left. \frac{g^{\mu \nu} \partial_\mu K^2 \partial_\nu K^2}{- 4 K^2} \right|_{r= r_+}} = \frac{r_+ - m}{a^2 + r_+^2} = \frac{r_+^2 - a^2}{2 r_+ \left(a^2 + r_+^2 \right)}
\ees 
(we denote the surface gravity with a tilde to distinguish it from the Newman-Penrose symbol $\kappa$). Since a Killing horizon is a just special type of isolated horizon, we shall extract the Kerr isolated horizon data from the Kerr Killing horizon and so we can make direct comparisons. \\
\\
The metric on the 2-dimensional horizon cross-section is simply 
\bes
\gamma = \left(r_+^2 + a^2 \cos^2 \theta \right) \td \theta^2 + \frac{\left( a^2 + r_+^2 \right)^2 \sin^2 \theta}{r_+^2 + a^2 \cos^2 \theta} \td \phi^2 \; . 
\ees
We shall choose $\xi\zo$ adapted to the angles $\{\theta , \phi \}$:
\bes
\xi^{\theta(0)} = \frac{1}{\sqrt{2} (r_+ + i a \cos \theta )}  \quad \xi^{\phi(0)} = \frac{ (a \cos \theta +i r_+)}{\sqrt{2} \sin \theta \left(a^2+r_+^2\right)} \; , 
\ees
and as one-forms they are
\bes
\xi_\theta\zo = \frac{r_+ -i a \cos \theta }{\sqrt{2}}  \quad \xi_{\phi}\zo = \frac{\left(a^2+r_+^2\right) \sin \theta}{\sqrt{2} (a \cos \theta -i r_+)} \; . 
\ees
so that $\bar{\xi}\zo \cdot \xi\zo = 1$ and $\xi\zo \cdot \xi\zo = \bar{\xi}\zo \cdot \bar{\xi}\zo=0$. From this we can derive the horizon connection $A =  \alpha\zo - \bar{\beta}\zo$. The last piece of horizon datum we need is $\pi\zo$. This can be extracted from the rotation 1-form $\omega^{(\ell)}$ on the horizon, which is given in \cite{Dobkowski-Rylko:2018ahh}. With our choice of parametrisation, the components read
\bes
\omega^{(\ell)}_\phi = \frac{a \sin ^2\theta  \left[a^2 \left(a^2-r_+^2\right) \cos^2 \theta- r_+^2 \left(a^2+3 r_+^2\right)\right]}{2 r_+ \left(a^2 \cos^2 \theta + r_+^2 \right)^2}  \; , \quad \omega^{(\ell)}_\theta = 0 \; . 
\ees
Then according to the relation
\bes
\omega^{(\ell)} = \pi\zo \xi\zo + \bar{\pi}\zo \bar{\xi}\zo - \tilde{\kappa} \td v \; , 
\ees
which holds on any isolated horizon, we obtain 
\bes
\pi\zo = \frac{a \sin \theta  (a \cos \theta -i r_+) \left[a^2 \left(a^2 \cos ^2 \theta - r_+^2 \left(1 + \cos^2 \theta \right)\right)-3 r_+^4\right]}{2 \sqrt{2} r_+ \left(a^2+r_+^2\right) \left(a^2 \cos^2 \theta + r_+^2\right)^2} \; . 
\ees
Because $r_+^2 + a^2 - 2mr_+ =0 $, we can use $m$ and $r_+$ interchangeably with $m = \frac{a^2 + r_+^2}{2 r_+}$. 
\subsection{Kerr horizon frame field functions}
Recall from section \ref{LOmetricfn}, these data are enough to find the metric functions to the lowest order. A short calculation reveals
\beas
U\fo &=&  \frac{r_+^2 - a^2}{2 r_+ \left(a^2 + r_+^2 \right)} \\
X^{\theta(1)} &=& 0 \\
X^{\phi(1)} &=& \frac{a^5-3 a^3 r_+^2+a^3 \left(a^2-r_+^2\right) \cos 2 \theta-6 a r_+^4}{2 r_+ \left(a^2+r_+^2\right)^2 \left(a^2 \cos 2 \theta+a^2+2 r_+^2\right)} \\
Z_\theta\fo &=& 0 \\
Z_\phi\fo &=& -\frac{a \sin ^2\theta \left[a^4+a^2 \left(a^2-r_+^2\right) \cos 2 \theta-3 a^2 r_+^2-6 r_+^4\right]}{r_+ \left(a^2 \cos 2 \theta+a^2+2 r_+^2\right)^2} \; . 
\eeas
We also have 
\beas
\xi^{\theta(1)} &=& -\frac{\left(a^2+r_+^2\right) \left(3 a^4-i a^3 r_+ \cos \theta+i a^3 r_+ \cos 3 \theta +3 a^2 \left(a^2+r_+^2\right) \cos 2 \theta+a^2 r_+^2-2 r_+^4\right)}{4 \sqrt{2} (m-r_+) (r_+-i a \cos \theta)^3 (r_++i a \cos \theta)^4} \\
\xi^{\phi(1)} &=& \frac{1}{32 \sqrt{2} r_+^2 \left(a^2+r_+^2\right)^2 (m-r_+) (r_++i a \cos \theta)^3}  \left\{ 4 a r_+ \left(a^6+22 a^4 r_+^2+17 a^2 r_+^4-8 r_+^6\right) \cot \theta
 \right. \\
&& 
-i \left(a^2-r_+^2\right) \csc \theta \left[a^6 \cos 4 \theta-a^6-a^4 r_+^2 \cos 4 \theta-27 a^4 r_+^2-20 a^2 r_+^4 \right.  \\
&& \left. \left.  -4 a^2 r_+^2 \left(5 a^2+3 r_+^2\right) \cos 2 \theta-4 i a^3 r_+ \left(a^2-r_+^2\right) \cos 3 \theta +16 r_+^6\right] \right\} \; ; 
\eeas
and as 1-form, such that the relations $\xi^{A(1)} \xi_A\zo + \xi^{A(0)} \xi_A\fo = \bar{\xi}^{A(1)} {\xi}_A\zo + \bar{\xi}^{A(0)} \xi_A\fo =0$ (the second equation is due to the definitions $\bar{\xi}^A \xi_A \equiv 1 $ and $\bar{\xi}^{A(0)} \xi_A\zo \equiv 1 $) hold:
\beas
\xi_\theta\fo &=& \frac{\left(a^2+r_+^2\right) \left(3 a^4-i a^3 r_+ \cos \theta +i a^3 r_+ \cos 3 \theta +3 a^2 \left(a^2+r_+^2\right) \cos 2 \theta +a^2 r_+^2-2 r_+^4\right)}{4 \sqrt{2} (m-r_+) (r_+-i a \cos \theta )^2 (r_++i a \cos \theta )^3} \\
\xi_\phi\fo &=& \frac{\sin \theta}{32 \sqrt{2} r_+^2 (m-r_+) (r_+-i a \cos \theta ) (r_++i a \cos \theta )^4} \left\{-4 a r_+ \left(a^6+22 a^4 r_+^2+17 a^2 r_+^4-8 r_+^6\right) \cos \theta  \right. 
\\
&&  +i \left(a^2-r_+^2\right) \left[a^6 \cos 4 \theta -a^6-a^4 r_+^2 \cos 4 \theta -27 a^4 r_+^2-20 a^2 r_+^4-4 a^2 r_+^2 \left(5 a^2+3 r_+^2\right) \cos 2 \theta \right.  \\
&& \left. \left. -4 i a^3 r_+ \left(a^2-r_+^2\right) \cos 3 \theta +16 r_+^6\right]\right\}
\eeas
\subsection{Derived horizon quantities}
The rest of the NP quantities on the horizon can be computed algebraically from $\tilde{k}$, $\xi\zo$ and $\pi\zo$ with the equations given in section \ref{indepHD}. They are, in computational order,
\beas
\epsilon\zo &=& \frac{r_+-m}{2 \left(a^2+r_+^2\right)} \\
\Psi_2\zo &=& -\frac{a^2+r_+^2}{2 r_+ (r_+-i a \cos \theta)^3}\\
\alpha\zo &=& -\frac{i \sin \theta \left(a^5-7 a^3 r_+^2-4 i r_+ \left(a^2+r_+^2\right)^2 \cot \theta \csc \theta+a^3 \left(a^2-r_+^2\right) \cos 2 \theta-10 a r_+^4\right)}{8 \sqrt{2} r_+ \left(a^2+r_+^2\right) (r_+-i a \cos \theta)^2 (r_++i a \cos \theta)} \\
\beta\zo &=& -\frac{i \csc \theta \left(-4 a \left(a^2-r_+^2\right) \sin ^2\theta \left(a^2 \cos 2 \theta+a^2+2 r_+^2\right)+16 i r_+ \left(a^2+r_+^2\right)^2 \cos \theta\right)}{32 \sqrt{2} r_+ \left(a^2+r_+^2\right) (r_+-i a \cos \theta) (r_++i a \cos \theta)^2} \\
\lambda\zo &=& \frac{a^2 \sin ^2\theta \left[\left(a^4+a^2 \left(a^2-r_+^2\right) \cos 2 \theta-3 a^2 r_+^2-6 r_+^4\right)^2-32 i a r_+^3 \left(a^2+r_+^2\right)^2 \cos \theta\right]}{32 r_+^2 \left(a^2+r_+^2\right) (m-r_+) (r_+-i a \cos \theta)^4 (r_++i a \cos \theta)^2} \\
\mu\zo &=& \frac{1}{256 r_+^2 \left(a^2+r_+^2\right) (m-r_+) \left(a^2 \cos ^2\theta+r_+^2\right)^3}\left[a^{10} \cos 6 \theta-2 a^{10}-2 a^8 r_+^2 \cos 6 \theta-180 a^8 r_+^2 \right.  \\
& & +a^6 r_+^4 \cos 6 \theta-450 a^6 r_+^4-312 a^4 r_+^6+48 a^2 r_+^8+2 a^4 \left(a^6-6 a^4 r_+^2-7 a^2 r_+^4+12 r_+^6\right) \cos 4 \theta \\
& & \left. -a^2 \left(a^8+190 a^6 r_+^2+561 a^4 r_+^4+480 a^2 r_+^6+48 r_+^8\right) \cos 2 \theta+128 r_+^{10} \right]\\
\Psi_3\zo &=& -\frac{3 i a \left(a^2+r_+^2\right) \sin \theta \left[\left(a^2-r_+^2\right) \left(a^2 \cos 2 \theta+a^2+2 r_+^2\right)+4 i a r_+ \left(a^2+r_+^2\right) \cos \theta\right]}{8 \sqrt{2} r_+^2 (m-r_+) (r_+-i a \cos \theta)^5 (r_++i a \cos \theta)} \\
\Psi_4\zo &=& \frac{3 a^2 \left(a^2+r_+^2\right) \sin ^2\theta \left[\left(a^2-r_+^2\right) \left(a^2 \cos 2 \theta+a^2+2 r_+^2\right)+4 i a r_+ \left(a^2+r_+^2\right) \cos \theta\right]^2}{32 r_+^3 (m-r_+)^2 (r_+-i a \cos \theta)^7 (r_++i a \cos \theta)^2} \; . 
\eeas
It can be checked explicitly that the relation 
\bes
2 \Psi_3^{(0)2} - 3 \Psi_2\zo \Psi_4\zo =0
\ees
which holds for any Petrov type D isolated horizon is satisfied. 

\subsection{Kerr first order quantities}
We can then use the radial equations \eqref{radialfoeqn} to obtain, for example
\beas
\rho\fo &=& -\frac{a^2+r_+^2}{2 r_+ (r_+-i a \cos \theta)^3} \\
\epsilon\fo &=& \frac{-1}{64 r_+^2 \left(a^2+r_+^2\right)^2 (r_+-i a \cos \theta)^3 (r_++i a \cos \theta)} \left[ 
a^8 (-\cos 4 \theta)+a^8-4 i a^7 r_+ \cos 3 \theta\right. \\
&&  +2 a^6 r_+^2 \cos 4 \theta+ 42 a^6 r_+^2  +8 i a^5 r_+^3 \cos 3 \theta-a^4 r_+^4 \cos 4 \theta+41 a^4 r_+^4-4 i a^3 r_+^5 \cos 3 \theta +32 r_+^8 \\
&& \left. +12 a^2 r_+^6+4 a^2 r_+^2 \left(5 a^4-2 a^2 r_+^2-3 r_+^4\right) \cos 2 \theta+4 i a r_+ \left(5 a^6+34 a^4 r_+^2+29 a^2 r_+^4-4 r_+^6\right) \cos \theta \right]
\\
\Psi_1\fo &=& -\frac{3 i a \left(a^2+r_+^2\right) \sin \theta}{2 \sqrt{2} r_+ (r_+-i a \cos \theta)^4 (r_++i a \cos \theta)} \; , 
\eeas
with the rest being too complicated so we omit them. 
\subsection{Next to lowest order frame field functions} 
These are given explicitly by
\beas
U\so &=&  {\pi}\zo {\Omega}\fo + \epsilon\fo + c.c. \\
X^{A(2)} &=& \pi\zo \xi^{A(1)} + \pi\fo \xi^{A(0)} + c.c. \\
\xi^{A(0)} Z_A\so &=& - \bar{\pi}\fo - \xi^{A(1)} Z_A\fo \\
\xi^{A(2)} &=& \mu\fo \xi^{A(0)} + \mu\zo \xi^{A(1)} + \bar{\lambda}\fo \bar{\xi}^{A(0)}+ \bar{\lambda}\zo \bar{\xi}^{A(1)} \; , 
\eeas
which can be computed to give
\beas
U\so &=& \frac{1}{8 r_+^2 \left(a^2+r_+^2\right)^2 \left(a^2 \cos 2 \theta+a^2+2 r_+^2\right)^3}  \left[ a^{10} (-\cos 6 \theta)+2 a^{10}+2 a^8 r_+^2 \cos 6 \theta+84 a^8 r_+^2 \right. \\
&& -a^6 r_+^4 \cos 6 \theta+226 a^6 r_+^4+216 a^4 r_+^6+48 a^2 r_+^8-2 a^4 \left(a^6-6 a^4 r_+^2-7 a^2 r_+^4+12 r_+^6\right) \cos 4 \theta \\
&& \left.  +a^2 \left(a^8+94 a^6 r_+^2+273 a^4 r_+^4+192 a^2 r_+^6-48 r_+^8\right) \cos 2 \theta-64 r_+^{10} \right] \\
X^{\theta(2)} &=& \frac{-a^2 \sin 2 \theta }{64 r_+^2 (m-r_+) \left(a^2 \cos ^2\theta +r_+^2\right)^5} \left[ 9 a^8+44 a^6 r_+^2+35 a^4 r_+^4+48 a^2 r_+^6 \right.  \\
&& \left.  +a^4 \left(3 a^4+4 a^2 r_+^2-7 r_+^4\right) \cos 4 \theta +4 a^2 \left(3 a^6+12 a^4 r_+^2+29 a^2 r_+^4+12 r_+^6\right) \cos 2 \theta +24 r_+^8\right] \\
X^{\phi(2)} &=& \frac{-a}{512 r_+^3 \left(a^2+r_+^2\right)^3 (m-r_+) \left(a^2 \cos ^2\theta +r_+^2\right)^4} \left[-4 a^{14} \cos 6 \theta -a^{14} \cos 8 \theta +5 a^{14}\right. \\
&&  +24 a^{12} r_+^2 \cos 6 \theta   +3 a^{12} r_+^2 \cos 8 \theta +393 a^{12} r_+^2-3 a^{10} r_+^4 \cos 8 \theta +1351 a^{10} r_+^4  +1152 r_+^{14} \\
&&  -56 a^8 r_+^6 \cos 6 \theta +a^8 r_+^6 \cos 8 \theta +1251 a^8 r_+^6+36 a^6 r_+^8 \cos 6 \theta -24 a^6 r_+^8+160 a^4 r_+^{10} \\
&&  +1344 a^2 r_+^{12}-4 a^4 \left(a^{10}-45 a^8 r_+^2-15 a^6 r_+^4+233 a^4 r_+^6+234 a^2 r_+^8-24 r_+^{10}\right) \cos 4 \theta \\
&& \left.  +4 a^2 \left(a^{12}+138 a^{10} r_+^2+352 a^8 r_+^4+158 a^6 r_+^6-153 a^4 r_+^8+160 a^2 r_+^{10}+368 r_+^{12}\right) \cos 2 \theta \right] \\
Z_\theta\so &=& \frac{ - 3 a^2 \left(a^2+r_+^2\right) \sin 2 \theta }{4 r_+^2 (r_+-m) \left(a^2 \cos 2 \theta+a^2+2 r_+^2\right)^4}  \left[3 a^6+13 a^4 r_+^2-8 a^2 r_+^4+a^4 \left(a^2-r_+^2\right) \cos 4 \theta\right. \\
&& \left. +4 a^2 \left(a^4+3 a^2 r_+^2+8 r_+^4\right) \cos 2 \theta+8 r_+^6\right] \\
Z_\phi\so &=& \frac{- 3 a \left(a^2+r_+^2\right)^2 \sin ^2\theta }{r_+ (r_+-m) \left(a^2 \cos 2 \theta+a^2+2 r_+^2\right)^5} \left[3 a^6+17 a^4 r_+^2+8 a^2 r_+^4+a^4 \left(a^2-5 r_+^2\right) \cos 4 \theta \right. \\
&& \left. +4 a^2 \left(a^4+3 a^2 r_+^2+4 r_+^4\right) \cos 2 \theta+8 r_+^6\right]
\eeas
\subsection{Kerr metric in Bondi-like coordinates}
Therefore the metric functions for Kerr, to the next to lowest order, are given by
\beas
g_{vv} &=& \frac{r \left(a^2-r_+^2\right)}{2 a^2 r_++2 r_+^3}-\frac{2 r^2 \left(a^2+r_+^2\right) \left(3 a^2 \cos 2 \theta +3 a^2-2 r_+^2\right)}{\left(a^2 \cos 2 \theta +a^2+2 r_+^2\right)^3}  + \mathcal{O}(r^3)\\
g_{v\theta} &=& \frac{-a^2 r^2 \sin 2 \theta }{32 \left(r_+^3-a^2 r_+\right) \left(a^2 \cos ^2\theta +r_+^2\right)^4} \left[ 9 a^8+50 a^6 r_+^2+5 a^4 r_+^4-24 a^2 r_+^6 \right. 
\\
&& \left. +a^4 \left(3 a^4-2 a^2 r_+^2-r_+^4\right) \cos 4 \theta +4 a^2 \left(3 a^6+12 a^4 r_+^2+35 a^2 r_+^4+30 r_+^6\right) \cos 2 \theta +24 r_+^8\right]   + \mathcal{O}(r^3) \\
g_{v\phi} &=& -\frac{2 a r \sin ^2\theta  \left[a^4+a^2 \left(a^2-r_+^2\right) \cos 2 \theta -3 a^2 r_+^2-6 r_+^4\right]}{r_+ \left(a^2 \cos 2 \theta +a^2+2 r_+^2\right)^2} \\
&& + \frac{a^3 r^2 \sin ^2\theta  \left(a^2 \cos 2 \theta +a^2+2 r_+^2\right)^5}{16384 r_+^2 \left(a^4-r_+^4\right) \left(a^2 \cos ^2\theta +r_+^2\right)^{10}} \left[4 a^{12} \cos 6 \theta +a^{12} \cos 8 \theta -5 a^{12}-24 a^{10} r_+^2 \cos 6 \theta \right. 
 \\
&& -3 a^{10} r_+^2 \cos 8 \theta +39 a^{10} r_+^2+3 a^8 r_+^4 \cos 8 \theta +2393 a^8 r_+^4+56 a^6 r_+^6 \cos 6 \theta -a^6 r_+^6 \cos 8 \theta \\
&&  +8541 a^6 r_+^6-36 a^4 r_+^8 \cos 6 \theta +12408 a^4 r_+^8+9200 a^2 r_+^{10}+4 a^2 \left(a^{10}-9 a^8 r_+^2-87 a^6 r_+^4 \right.  \\
&& \left. -199 a^4 r_+^6-270 a^2 r_+^8-204 r_+^{10}\right) \cos 4 \theta +4 \left(-a^{12}+6 a^{10} r_+^2+512 a^8 r_+^4+2146 a^6 r_+^6 \right.  \\
&&  \left. \left.  +3321 a^4 r_+^8+2000 a^2 r_+^{10}+208 r_+^{12}\right) \cos 2 \theta +3264 r_+^{12}\right]  + \mathcal{O}(r^3) \\
g_{\theta \theta} &=& a^2 +r_+^2 \cos ^2\theta +\frac{r r_+ \left(a^2+r_+^2\right)^2 \left(3 a^2 \cos 2 \theta +3 a^2-2 r_+^2\right)}{\left(a^2-r_+^2\right) \left(a^2 \cos ^2\theta +r_+^2\right)^2}  + \mathcal{O}(r^2) \\
g_{\phi \phi} &=& \frac{2 \left(a^2+r_+^2\right)^2 \sin ^2\theta }{a^2 \cos 2 \theta +a^2+2 r_+^2}-\frac{r \left(a^2+r_+^2\right) \sin ^2\theta }{32 \left(r_+^3-a^2 r_+\right) \left(a^2 \cos ^2\theta +r_+^2\right)^4} \left[a^{10} (-\cos 6 \theta )+2 a^{10}+2 a^8 r_+^2 \cos 6 \theta \right. \\
&& +84 a^8 r_+^2 -a^6 r_+^4 \cos 6 \theta +226 a^6 r_+^4+216 a^4 r_+^6+48 a^2 r_+^8-2 a^4  \cos 4 \theta \left(a^6-6 a^4 r_+^2-7 a^2 r_+^4 \right.  \\
&& \left. \left. +12 r_+^6\right)  +a^2 \left(a^8+94 a^6 r_+^2+273 a^4 r_+^4+192 a^2 r_+^6-48 r_+^8\right) \cos 2 \theta -64 r_+^{10}\right]  + \mathcal{O}(r^2) \\
g_{\theta \phi} &=& -\frac{32 a^3 r r_+^2 \left(a^2+r_+^2\right)^2 \sin ^3\theta  \cos \theta }{\left(a^2-r_+^2\right) \left(a^2 \cos (2 \theta )+a^2+2 r_+^2\right)^3}  + \mathcal{O}(r^2)\; . 
\eeas 
In the extremal limit where a $\rightarrow r_+$ the $\mathcal{O}(r)$ term in $g_{vv}$, which is proportional to the surface gravity, vanishes as expected. However, one cannot obtain the extremal Kerr metric by taking this limit as there would be $1/(a^2 - r_+^2)$ singularity everywhere.  
\section{Metric near Kerr-dS isolated horizon}
The Kerr-dS metric in Boyer-Lindquist coordinates is given by 
\bes
\td s^2 = -\frac{\Delta_r}{\rho^2} \left( \td t - \frac{a \sin^2 \theta}{\Xi} \td \phi \right)^2 + \frac{\rho^2}{\Delta_r} \td r^2 + \frac{\rho^2}{\Delta_\theta} \td \theta^2 + \frac{\Delta_\theta \sin^2 \theta}{\rho^2} \left( a \td t - \frac{r^2+a^2}{\Xi} \td \phi \right)^2 
\ees
where 
\bes
\rho^2 = r^2 + a^2 \cos^2 \theta \; , \quad \Xi= 1- \frac{a^2}{\ell^2} \; , \quad \Delta_\theta = 1 - \frac{a^2}{\ell^2}\cos^2 \theta \quad \mathrm{and} \quad \Delta_r = (r^2 + a^2)\left(1 - \frac{r^2}{\ell^2} \right) - 2mr \; ; 
\ees
$m$ is the mass and $a$ is the angular momentum and $\ell \equiv \tfrac{1}{g} = \sqrt{\tfrac{3}{\Lambda}}$ is the radius of dS. The event horizon is located at the $r= r_+$, the larger root of $ \Delta_r$, which is a Killing horizon generated by the Killing field
\bes
K = \frac{\partial}{\partial t} +\frac{a \left(1 + a^2 g^2 \right) }{r_+^2 + a^2} \frac{\partial}{\partial \phi}\; , 
\ees
with surface gravity 
\bes
\tilde{\kappa} = \frac{r_+^2( 1- 3 g^2 r_+^2) - a^2 (1+ g^2 r_+^2) }{2 r_+ (a^2 + r_+^2)} \; . 
\ees 
In order to describe a regular spacetime with well separated inner $(r=r_-)$ and outer $(r=r_+)$ horizons sandwiched between a pair of cosmological horizons $(r=\pm \ell)$, it is necessarily that $r_+< \ell$. We also assume that rotation is slow and the cosmological constant is small, i.e. $a\ll r_+$ and $g\ll r_+$. We have also chosen the above parametrisation so that we recover Kerr as shown in the previous section in the limit of $g \rightarrow 0$. Because $\left(r_+^2 + a^2 \right) \left( 1- r_+^2 g^2  \right) - 2mr_+ =0 $, we can use $m$ and $r_+$ interchangeably with $m = \frac{\left( a^2 + r_+^2 \right) \left( 1- r_+^2 g^2 \right)}{2 r_+}$.   \\
\\
The metric on the 2-dimensional horizon cross-section is simply 
\bes
\gamma = \frac{r_+^2 + a^2 \cos^2 \theta }{1+ a^2 g^2 \cos^2 \theta } \td \theta^2 + \frac{\left( a^2 + r_+^2 \right)^2 \left(1+ a^2 g^2 \cos^2 \theta \right)  \sin^2 \theta}{ \left( 1 + a^2 g^2 \right)^2 \left(r_+^2 + a^2 \cos^2 \theta \right)} \td \phi^2 \; . 
\ees
A convenient choice of the frame field $\xi\zo$ adapted to the angles $\{\theta , \phi \}$ is:
\bes
\xi^{\theta(0)} = \frac{\sqrt{1 + a^2g^2 \cos^2 \theta}}{\sqrt{2} (r_+ + i a \cos \theta )}  \quad \xi^{\phi(0)} = \frac{ (1+a^2 g^2) (a \cos \theta +i r_+)}{\sqrt{2} \sin \theta \left(a^2+r_+^2\right)\sqrt{1+ a^2 g^2 \cos^2 \theta}} \; , 
\ees
and as one-forms they are
\bes
\xi_\theta\zo = \frac{r_+ -i a \cos \theta }{\sqrt{2}\sqrt{1+a^2 g^2 \cos^2 \theta} }  \quad \xi_{\phi}\zo = \frac{\left(a^2+r_+^2\right)\sqrt{1+a^2 g^2 \cos^2 \theta} \sin \theta}{\sqrt{2} (1+a^2 g^2)(a \cos \theta -i r_+)} \; . 
\ees
so that $\bar{\xi}\zo \cdot \xi\zo = 1$ and $\xi\zo \cdot \xi\zo = \bar{\xi}\zo \cdot \bar{\xi}\zo=0$. The last piece of information reqired is $\pi\zo$. This can be extracted from the rotation 1-form $\omega^{(\ell)}$ on the horizon, which is also given in \cite{Dobkowski-Rylko:2018ahh}. With our choice of parametrisation, the components read
\bes
\omega^{(\ell)}_\phi = \frac{a (1- r_+^2 g^2) \sin ^2\theta  \left[a^2 \left(a^2-r_+^2\right) \cos^2 \theta- r_+^2 \left(a^2+3 r_+^2\right)\right]}{2 r_+ \left( 1+a^2 g^2 \right) \left(a^2 \cos^2 \theta + r_+^2 \right)^2}  \; , \quad \omega^{(\ell)}_\theta = 0 \; . 
\ees
Then according to the relation
\bes
\omega^{(\ell)} = \pi\zo \xi\zo + \bar{\pi}\zo \bar{\xi}\zo - \tilde{\kappa} \td v \; , 
\ees
which holds on any isolated horizon, we obtain 
\bes
\pi\zo = \frac{a \sin \theta \left( 1- r_+^2 g^2 \right)  (a \cos \theta -i r_+) \left[a^2 \left(a^2 \cos ^2 \theta - r_+^2 \left(1 + \cos^2 \theta \right)\right)-3 r_+^4\right]}{2 \sqrt{2} r_+ \left(a^2+r_+^2\right) \left(a^2 \cos^2 \theta + r_+^2\right)^2 \sqrt{1+a^2 g^2 \cos^2 \theta} } \; . 
\ees
\subsection{Kerr-dS horizon geometry}
We already have enough initial data to compute the frame field functions to the lowest order. A short calculation reveals 
\beas
U\fo &=& \frac{r_+^2( 1- 3 g^2 r_+^2) - a^2 (1+ g^2 r_+^2) }{2 r_+ (a^2 + r_+^2)} \\
X^{\theta(1)} &=& 0 \\
X^{\phi(1)} &=& \frac{a (1+ a^2 g^2) (1-r_+^2 g^2 ) \left[ a^4 \cos^2 \theta - a^2 r_+^2 \left( 1+ \cos^2 \theta \right) - 3 r_+^4 \right]}{2 r_+ (a^2 + r_+^2)^2 (r_+^2 + a^2 \cos^2 \theta) (1+ a^2 g^2 \cos^2\theta)} \\
Z\fo_\theta &=& 0 \\
Z\fo_\phi &=& - \frac{a (1- g^2 r_+^2)\left( a^4 \cos^2 \theta - a^2 r_+^2 (1 + \cos^2 \theta) - 3 r_+^4 \right) \sin^2 \theta }{2 r_+ \left( 1+ a^2 g^2 \right) \left( r_+^2 + a^2 \cos^2 \theta \right)^2 } \; . 
\eeas
Therefore the remaining data to specify the horizon geometry of Kerr-dS in Gaussian null coordinates are given by 
\beas
g_{vv} &=& \frac{r \left(a^2 \left(1 + r_+^2 g^2 \right) + r_+^2 \left( 3 r_+^2 g^2 -1 \right) \right) }{2 r_+ \left( a^2 + r_+^2  \right)} \\
g_{v\theta} &=& 0 \\
g_{v\phi} &=& - \frac{r a \sin^2 \theta \left(1- r_+^2 g^2 \right)\left( a^4 \cos^2 \theta - a^2 r_+^2 (1 + \cos^2 \theta) - 3 r_+^4 \right) }{r_+ \left( 1 + a^2 g^2 \right) \left(r_+^2 + a^2 \cos^2 \theta \right)^2} \; . 
\eeas

\section{Summary and outlook}
In this paper, we considered spacetime in the neighbourhood of an isolated horizon in 4-dimensions. We showed that if the horizon symmetry extends to the bulk, then the spacetime geometry is completely determined order by order in the radial direction by the connection and the NP spin coefficient $\pi$ specified on the horizon cross-section. Using this fact, we computed explicitly the Kerr metric up to first order in Bondi-like coordinates. We also gave explicitly the horizon metric of Kerr-dS in Bondi-like coordinates. It still remains an open problem if the Kerr metric can be expressed in Bondi-like coordinates in terms of fundamental functions.

\section*{Acknowledgement }
This work was supported by the Polish National Science Centre grant No. 2015/17/B/ST2/02871.

\bibliography{KerrIH}
\bibliographystyle{ieeetr}

\end{document}